\documentclass[preprint,showpacs,preprintnumbers,amsmath,amssymb]{revtex4}

\usepackage{graphicx}% Include figure files
\usepackage{dcolumn}% Align table columns on decimal point
\usepackage{bm}% bold math

\begin{document}

%\preprint{APS/123-QED}

\title{Magnetic properties of PrPd$_{2}$Si$_{2}$ and PrPt$_{2}$Si$_{2}$}
\author{V. K. Anand}
\email{vivekkranand@gmail.com}
\author{Z. Hossain}
\affiliation{ Department of Physics, Indian Institute of Technology, Kanpur 208016, India}
\author{C. Geibel}
\affiliation{Max Planck Institute for Chemical Physics of Solids, 01187 Dresden, Germany}
%\email{ }
%\homepage{}
\date{\today}% It is always \today, today,
             %  but any date may be explicitly specified
             
\begin{abstract}

We have investigated the two rare-earth intermetallic compounds PrPd$_{2}$Si$_{2}$ and PrPt$_{2}$Si$_{2}$ by means of magnetization, electrical resistivity and heat capacity measurements. While PrPd$_{2}$Si$_{2}$ exhibits an antiferromagnetic ordering at 3 K, no magnetic ordering is observed in PrPt$_{2}$Si$_{2}$ down to 2 K. The different magnetic behaviors of these two compounnds are due to different crystalline electric field (CEF) level schemes. The specific heat data suggest a quasi-quartet ground state in PrPd$_2$Si$_2$ in contrast to a nonmagnetic singlet ground state in PrPt$_2$Si$_2$. This difference is attributed to the loss of a mirror plane upon changing the crystal structure from the ThCr$_2$Si$_2$ type (PrPd$_2$Si$_2$) to the CaBe$_2$Ge$_2$ type (PrPt$_2$Si$_2$). Further on, a large magnetoresistance is also observed in the magnetically ordered state of PrPd$_2$Si$_2$.           

\end{abstract}

% PACS, the Physics and Astronomy
                             % Classification Scheme.
%\keywords{Borocarbides, Crystal Field Effect, Strongly correlated electron system, Heavy fermions}%Use showkeys class option if keyword
                              %display desired
\maketitle

%\section{\label{}Introduction\protect \lowercase{} }

\section*{Introduction}

Pr-based intermetallic compounds have evolved as a topic of current interest among the condensed matter physicists as some of these compounds exhibit interesting physical properties. While in Ce-compounds the relative strengths of the RKKY and Kondo interactions decide  the ground state properties, in case of Pr-compounds the ground state depends critically on the crystal electric field (CEF) level scheme. For example, in PrOs$_{4}$Sb$_{12}$ a small CEF splitting energy of 0.7 meV and quadrupolar excitations lead to unconventional heavy-fermion superconductivity [1-3]. Pr$_{2}$Rh$_{3}$Ge$_{5}$ exhibits heavy fermion behaviour in which low lying crystal field excitations are responsible for the mass enhancement instead of the the usual Kondo effect \cite{4}. We have also investigated PrRh$_{2}$Si$_{2}$ in view of the unusual superconducting and magnetic properties of CeRh$_{2}$Si$_{2}$ and YbRh$_{2}$Si$_{2}$. We found an antiferromagnetic ordering at 68 K in PrRh$_{2}$Si$_{2}$ which is anomalously high compared to the expected de-Gennes-scaled transition temperature of 5.4 K \cite{5}. Further, we decided to investigate PrPd$_{2}$Si$_{2}$ and PrPt$_{2}$Si$_{2}$ in view of the interesting features of CePd$_{2}$Si$_{2}$ and CePt$_{2}$Si$_{2}$. While CePd$_{2}$Si$_{2}$ is a heavy-fermion antiferromagnet system which exhibits pressure induced superconductivity \cite{6}, CePt$_{2}$Si$_{2}$ is a Kondo lattice non-Fermi liquid system that does not order down to 60 mK \cite{7}. A preliminary magnetization study on PrPt$_{2}$Si$_{2}$ reports it to be paramagnetic down to 1.8 K \cite{8}. We report here our results of magnetization, electrical resistivity, and heat capacity studies of PrPd$_{2}$Si$_{2}$ and PrPt$_{2}$Si$_{2}$.

\section*{Experimental}

Polycrystalline samples of PrPd$_{2}$Si$_{2}$ and PrPt$_{2}$Si$_{2}$ and their La-analogs were prepared by standard arc-melting on a water cooled copper hearth under an inert argon atmosphere starting with high purity (99.99\% and above) elements in stoichiometric ratio. To ensure a proper mixing of the constituents, arc melted ingots were flipped and remelted several times. Weight loss during the melting process was less than 0.5\%. The samples were annealed at 1000 $^{o}$C for one week to improve the sample quality. Thereafter, the samples were characterized using powder X-ray diffraction and scanning electron microscopy (SEM) equipped with energy dispersive X-ray analysis (EDAX). Magnetization measurements were performed using a commercial SQUID magnetometer. The heat capacity was measured using relaxation method in a physical property measurement system (PPMS-Quantum Design). The electrical resistivity was measured by four probe ac technique using the ac transport option of PPMS.

\section*{Results and Discussion}

While the compounds PrPd$_{2}$Si$_{2}$, LaPd$_{2}$Si$_{2}$ crystallize in ThCr$_{2}$Si$_{2}$-type tetragonal structure (space-group {\it I4/mmm}), PrPt$_{2}$Si$_{2}$ and LaPt$_{2}$Si$_{2}$ form in CaBe$_{2}$Ge$_{2}$-type primitive tetragonal structure (space group {\it P4/nmm}). The lattice parameters and unit cell volumes for these compounds are listed in table 1. The lattice parameters obtained for PrPt$_{2}$Si$_{2}$, LaPt$_{2}$Si$_{2}$, and LaPd$_{2}$Si$_{2}$ are close to the values reported in references \cite{8} and \cite{9}. Though all the peaks in X-ray diffraction pattern of PrPd$_{2}$Si$_{2}$ and PrPt$_{2}$Si$_{2}$ are well indexed, the scanning electron micrographs show the presence of impurity phase(s) which we estimated to be less than 3\% in PrPd$_{2}$Si$_{2}$ and about 6\% in PrPt$_{2}$Si$_{2}$.  The EDAX analysis confirms the desired stoichiometry of 1:2:2.

\begin{table}
\caption{\label{tab:table1} Lattice parameters and unit cell volumes of tetragonal compounds RT$_{2}$Si$_{2}$ (R = La, Pr and T = Pd, Pt).}
\begin{ruledtabular}
\begin{tabular}{lcccc}

Compounds&$a$ (\AA)&$c$ (\AA) &$V$ (\AA$^3$) & Space group \\
\hline
PrPd$_{2}$Si$_{2}$& 4.2232(9) & 9.874(3)& 176.12(5) & {\it I4/mmm} \\
LaPd$_{2}$Si$_{2}$& 4.2835(1) & 9.862(6)& 180.94(1) & {\it I4/mmm} \\
PrPt$_{2}$Si$_{2}$& 4.2426(8) & 9.781(3)& 176.06(4) & {\it P4/nmm} \\
LaPt$_{2}$Si$_{2}$& 4.2824(1) & 9.827(6)& 180.21(0) & {\it P4/nmm} \\

\end{tabular}
\end{ruledtabular}
\end{table}

\section*{A. PrPd$_{2}$Si$_{2}$}

The magnetic susceptibility data of PrPd$_{2}$Si$_{2}$ are shown in figure 1. Magnetic susceptibility follows the Curie-Weiss behavior above 50 K. Fitting the inverse susceptibility data to the expression 1/$\chi$ = ($T - \theta_{p}$)/$C$, we obtained an effective moment $\mu_{eff}$ = 3.59 $\mu_{B}$ (theoretically expected value for Pr$^{3+}$ ions is 3.58 $\mu_{B}$) and a Weiss temperature $\theta_{p}$ = - 4.2 K.
The low temperature susceptibility data (shown in the inset of figure 1) show a peak at 3.5 K. The position of the peak shifts towards lower temperatures with increasing field, thereby confirming the antiferromagnetic nature of the transition at 3.5 K. The isothermal magnetization data at 2 K (figure 2) exhibits a very smooth metamagnetic-type transition at 4.8 T. The critical field for the metamagnetic transition is determined from the $dM/dB$ {\it vs.} $B$ plot  (inset of figure 2). No saturation is observed upto B = 6.0 T.

\begin{figure}
\centering
\includegraphics[width = 10 cm, keepaspectratio] {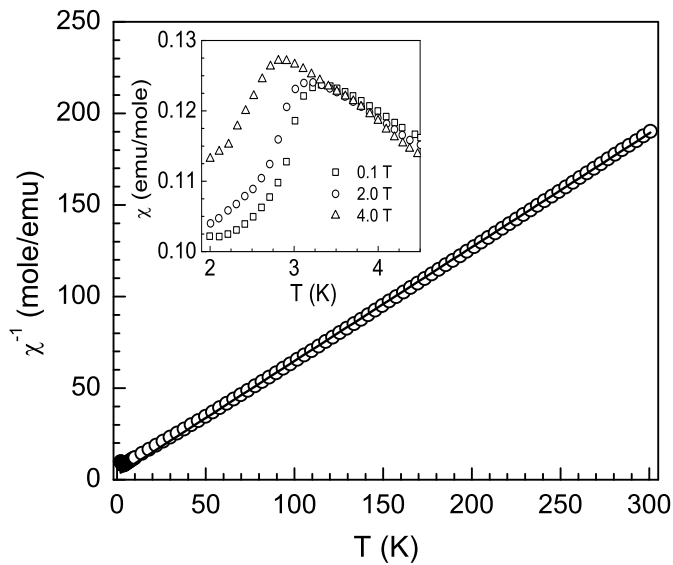}
\caption{\label{fig1} Inverse magnetic susceptibility plot of PrPd$_{2}$Si$_{2}$ in the temperature range 2 -- 300 K. The inset shows the low temperature susceptibility data at three different fields.}
\end{figure}

\begin{figure}
\centering
\includegraphics[width=10cm, keepaspectratio] {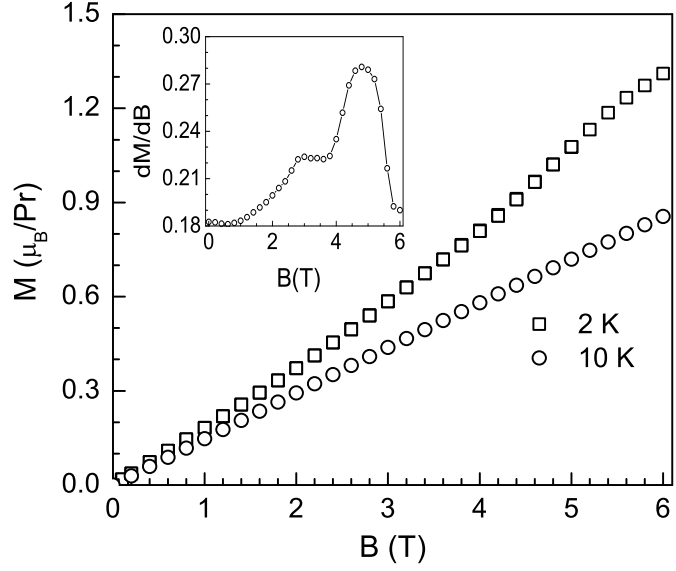}
\caption{\label{fig2} The magnetic field dependence of magnetization of PrPd$_{2}$Si$_{2}$ at 2 and 10 K. The inset shows the derivative of magnetization data at 2 K.}
\end{figure}

\begin{figure}
\centering
\includegraphics[width=10cm, keepaspectratio] {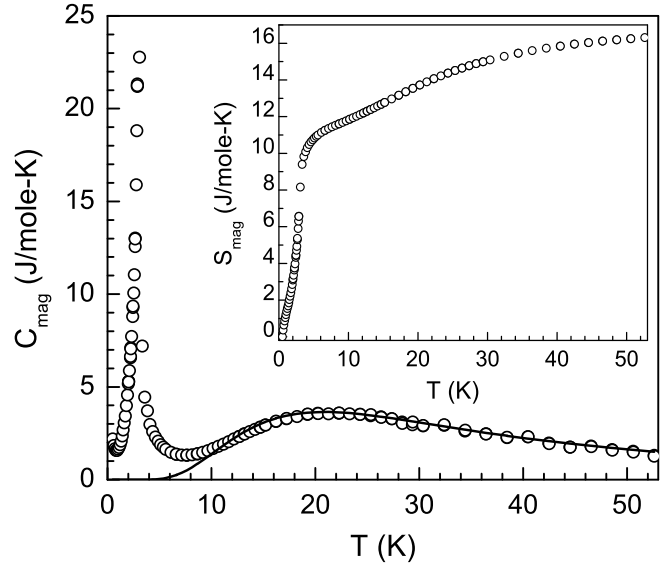}
\caption{\label{fig3} The magnetic contribution to the specific heat of PrPd$_{2}$Si$_{2}$ as a function of temperature in the temperature range 0.5 -- 52 K. The solid line represents the fit for equally degenerate two level Schottky anomaly with a separation of 50 K. The inset shows the temperature dependence of magnetic entropy.}
\end{figure}

Figure 3 shows the magnetic contribution to the specific heat of PrPd$_{2}$Si$_{2}$ which we obtained after subtracting the lattice contribution assuming it to be roughly equal to that of the nonmagnetic analog LaPd$_{2}$Si$_{2}$. The specific heat data of PrPd$_{2}$Si$_{2}$ exhibits a sharp $\lambda$-type peak at 3 K which confirms the intrinsic nature of magnetic order in this compound. The magnetic entropy reaches a value close to $Rln4$ (= 11.52 J/mole-K) at 7.5 K, suggesting a quasi-quartet ground state. We also observe a pronounced Schottky type anomaly with a broad maximum centered around 20 K which we attribute to the crystal field effect. The position of Schottky peak suggests that the excited states lie about 50 K above the low lying CEF states. The peak height of the Schottky anomaly is consistent with equal degeneracy between the low lying states and the excited states, suggesting the presence of four CEF levels close to 50 K. In a tetragonal symmetry one expects the CEF levels of Pr$^{3+}$ to split into five singlets and two doublets. Thus in PrPd$_{2}$Si$_{2}$, the temperature dependence of the entropy indicate a separation into four low lying levels (either two doublets, or one doublet and two singlets, or four singlets) separated by less than 10 K, four further levels around 50 K, and an upper singlet at much higher energy. Since in RPd$_{2}$Si$_{2}$ the CEF schemes seem to be determined by the higher order terms in the CEF hamiltonian \cite{10,11}, a simple preliminary guess of the CEF sheme of PrPd$_{2}$Si$_{2}$ can not be given.

\begin{figure}
\centering
\includegraphics[width=11cm, keepaspectratio] {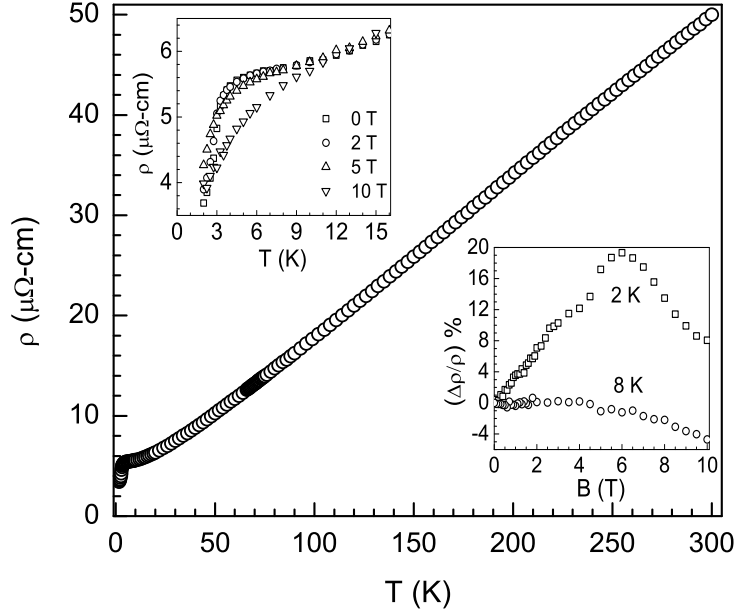}
\caption{\label{fig4} Temperature dependence of electrical resistivity of PrPd$_2$Si$_2$ in the temperature range 2 -- 300 K. The upper inset shows the low temperature resistivity data at different fields. The lower inset shows the magnetoresistance normalized to $\rho(T)$ at B = 0.}
\end{figure}

In the paramagnetic regime the electrical resistivity decreases almost linearly with decreasing temperature (figure 4) and merge into a constant value of 5.5 $\mu \Omega$-cm below 10 K. The resistivity drops rapidly below the ordering temperature due to reduction of spin disorder scattering. The resistivity at 2 K, where $\rho(T)$ is still decreasing with T, is 3.4 $\mu \Omega$-cm  leading to a lower bound of $\sim$ 15 for the residual resistivity ratio. This evidences a good quality of our polycrystalline sample. The upper inset of figure 4 shows the effect of a magnetic field on the resistivity. The resistivity anomaly related to the magnetic order smoothens out for B $>$ 2 T. The magnetoresistance $\Delta \rho$/$\rho$ = [$\rho$($B$)-$\rho$(0)]/$\rho$(0) is shown in the lower inset of figure 4. In the ordered state (at 2 K) the magnetoresistance initially increases with increasing field, peaks at 6 T and decreases for higher fields. Such a behavior of the magnetoresistance is expected for an antiferromagnetic state. For T $\ll$ T$_N$, an increasing magnetic field is first weakening the AF-state, leading to an increase of spin scattering, but above the metamagnetic transition the formation of the field aligned state results in a decrease of spin scattering. In the paramagnetic state the magnetoresistane is always negative, since alignment of the moments in the external field reduces the spin scattering. Thus, magnetic susceptibility, magnetization, specific heat and magnetoresistance data provide conclusive evidence for an antiferromagnetic state in PrPd$_2$Si$_2$ below 3.0 K.

\section*{B. PrPt$_{2}$Si$_{2}$}

Figure 5 shows the magnetic susceptibility data of PrPt$_{2}$Si$_{2}$. No anomaly is observed in susceptibility data down to 2 K implying the absence of magnetic ordering in this compound. The paramagnetic susceptibility displays a Curie-Weiss character. From a fit of the inverse susceptibility data above 50 K to the expression 1/$\chi$ = ($T - \theta_{p}$)/$C$ we found the effective moment to be $\mu_{eff}$ = 3.46 $\mu_{B}$ which is very close to the value of 3.58 $\mu_{B}$ expected for Pr$^{3+}$ ions. The Curie-Weiss temperature $\theta_{p}$ = +18.8 K indicates dominant ferromagnetic exchange in this compound. The isothermal magnetization curve is linear with field at 2 and 20 K which is consistent with a paramagnetic state down to 2 K in this compound.

\begin{figure}
\centering
\includegraphics[width=10cm, keepaspectratio] {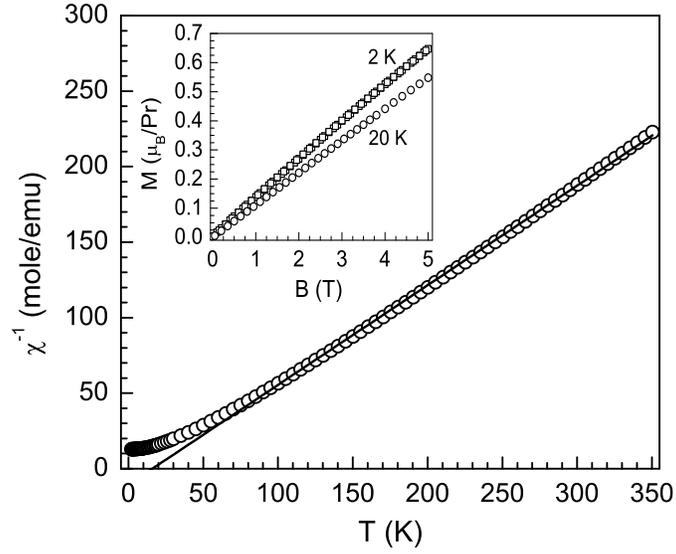}
\caption{\label{fig5} Inverse magnetic susceptibility plot of PrPt$_{2}$Si$_{2}$ at a field of 1.0 T. The inset shows magnetization as a function of field at two different temperatures of 2 and 20 K.}
\end{figure}

\begin{figure}
\centering
\includegraphics[width=10cm, keepaspectratio] {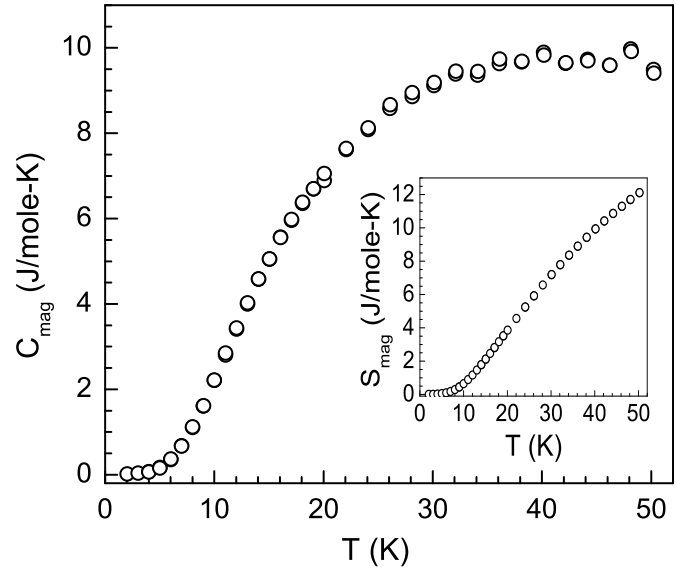}
\caption{\label{fig6} Temperature dependence of magnetic contribution to the specific heat of PrPt$_{2}$Si$_{2}$ in the temperature range (2 -- 50 K). The inset shows the temperature dependence of magnetic entropy.}
\end{figure}

The magnetic contribution to the specific heat of PrPt$_{2}$Si$_{2}$ (figure 6) was obtained by subtracting the specific heat data of LaPt$_{2}$Si$_{2}$ from that of PrPt$_{2}$Si$_{2}$. No signature of magnetic or superconducting transition is observed in the specific heat data, which is consistent with the paramagnetic ground state inferred from the magnetic susceptibility data. From the low temperature specific heat data we estimate a $\gamma$ value of $\sim$ 3~mJ/mole-K$^{2}$. Such a low magnitude of specific heat at low T rules out the possibility of magnetic ordering even at lower temperature. A broad Schottky type  anomaly is observed in the magnetic part of the specific heat above 20 K. The upturn in the magnetic contribution (C$_{mag}$) to the specific heat around 10 K could be reproduced with a first excited doublet separated from the singlet ground state by approximately 50 K. For the case of a singlet ground state and a doublet as first excited state one would expect a peak height of 6.2 J/mole-K in C$_{mag}$. However, the experimentally observed value is 9.2 J/mole-K, clearly indicating that further higher lying states are also contributing. The large separation between the singlet ground state and the first excited state is responsible for the absence of magnetic order in PrPt$_{2}$Si$_{2}$.

Usually the CEF schemes of homologous RT$_2$X$_2$ compounds with different T- (or X-) elements of one row in the periodic table are quite similar, because neither the effective ionic charge at the ligands nor their distance to the R-atoms changes significantly. Thus, in the present case, the strong differences between the CEF schemes of PrPd$_{2}$Si$_{2}$ and PrPt$_{2}$Si$_{2}$ are very likely related to the loss of a mirror plane at the Pr-site upon changing the crystal structure from the ThCr$_2$Si$_2$ to the CaBe$_2$Ge$_2$ type, although both structure types are closely related. This difference in the structures is likely also responsible for the change from dominant antiferromagnetic exchange in the Pd-based compound to dominant ferromagnetic exchange in the Pt-based compound.

\section*{Summary and Conclusions}

We have investigated magnetic properties of PrPd$_{2}$Si$_{2}$ and PrPt$_{2}$Si$_{2}$. From a detailed measurements of magnetization, specific heat, electrical resistivity and magnetoresistance we have established antiferromagnetic ordering in PrPd$_{2}$Si$_{2}$ below 3 K. Below T$_N$ this compound also exhibits field induced metamagnetic transition and large magnetoresistance (20\%) at a magnetic field of 6 Tesla. In contrast, no magnetic order is observed in PrPt$_{2}$Si$_{2}$ down to 2 K. We attribute this paramagnetic ground state in PrPt$_{2}$Si$_{2}$ to the CEF scheme with a singlet ground state and well separated first excited state located around 50 K. The pronounced differences between PrPd$_{2}$Si$_{2}$ and PrPt$_{2}$Si$_{2}$ are attributed to the loss of a mirror plane upon changing the crystal structure from the ThCr$_2$Si$_2$ to the CaBe$_2$Ge$_2$ type. Further investigations are needed to determine the exact magnetic structure of PrPd$_{2}$Si$_{2}$ and the crystal field level schemes in these two compounds.

\end{document}